\documentclass[a4paper]{quantumarticle}
\pdfoutput=1

\usepackage[utf8]{inputenc}
\usepackage[T1]{fontenc}
\usepackage[english]{babel}

\usepackage{amsmath}
\usepackage{amssymb}
\usepackage{braket}

\usepackage{xcolor}
\usepackage{orcidlink}

\usepackage{hyperref}

\hypersetup{
  colorlinks=true,
  linkcolor=blue,      
  citecolor=blue,      
  urlcolor=blue        
}

\usepackage{tikz}
\usepackage{lipsum}

\begin{document}

\title{A hardware-native time–frequency GKP logical qubit toward fault-tolerant photonic operation}

\author{Tai Hyun Yoon\,\orcidlink{0000-0002-2408-9295}}
\email[]{thyoon@korea.ac.kr}
\affiliation{Department of Physics and Center for Molecular Spectroscopy and Dynamics, IBS, Korea University, 145 Anam-ro, Seongbuk-gu, Seoul 02841, Republic of Korea}


\begin{abstract}
We realize a hardware-native time--frequency Gottesman--Kitaev--Preskill (GKP)
logical qubit encoded in the continuous phase space of single photons,
establishing a propagating photonic implementation of bosonic grid encoding.
Finite-energy grid states are generated deterministically using coherently
driven entangled nonlinear biphoton sources that produce single-photon
frequency-comb supermodes.
An optical-frequency-comb reference anchors the time--frequency phase space
and enforces commuting displacement stabilizers directly at the hardware level,
continuously defining the logical subspace.
Timing jitter, spectral drift, and phase noise map naturally onto
Gaussian displacement errors within this lattice, yielding intrinsic
correctability inside a stabilizer cell.
Logical operations correspond to experimentally accessible phase and delay
controls, enabling deterministic state preparation and manipulation.
Building on the modal time--frequency GKP framework, we identify a
concrete pathway toward active syndrome extraction and deterministic
displacement recovery using ancillary grid states and interferometric
time--frequency measurements.
These primitives establish a hardware-compatible route for integrating
the time--frequency GKP logical layer into erasure-aware and
fusion-based fault-tolerant photonic architectures.
\end{abstract}

\maketitle

\section{Introduction}
\label{sec:introduction}

Fault-tolerant quantum information processing requires physical encodings
that suppress dominant noise processes while remaining compatible with
realistic hardware constraints.
Bosonic encodings based on continuous degrees of freedom provide a powerful
route toward this goal by embedding logical qubits into structured subspaces
of an infinite-dimensional Hilbert space, where small errors correspond to
correctable displacements~\cite{Weedbrook2012,Albert2018,Terhal2020}.
Within this paradigm, the Gottesman--Kitaev--Preskill (GKP) framework
defines logical qubits stabilized by a lattice of commuting phase-space
displacement operators~\cite{Gottesman2001,Cai2021}.
Despite substantial theoretical and experimental progress across trapped ions,
superconducting cavities, and microwave platforms
\cite{Fluehmann2019,CampagneIbarcq2020,deNeeve2022,LachanceQuirion2024},
a central open challenge remains: to realize bosonic logical qubits in a
manner that is hardware-native, metrologically anchored, and directly
expressible in terms of experimentally accessible control operations,
rather than imposed abstractly at the level of idealized quadratures.

Here we establish a bosonic logical qubit encoded in the
\emph{time--frequency (TF) phase space of single photons},
providing a propagating photonic implementation of GKP encoding.
Time and frequency form a canonically conjugate pair admitting a natural
continuous-variable phase-space description,
with calibrated delays, phase shifts, and frequency translations
acting as physical displacement operators
\cite{Fabre2022,Descamps2023a}.
When stabilized by an external optical-frequency-comb reference,
this TF phase space acquires a discrete lattice structure
that supports finite-energy grid states directly analogous to GKP codewords
\cite{Gottesman2001,Descamps2024a}.
In contrast to idealized constructions,
the stabilizers of the logical subspace are enforced physically through
metrological reference locking, continuously anchoring the logical encoding
in laboratory time.

The physical realization is based on coherently driven
entangled nonlinear biphoton sources (ENBSs),
which generate single-photon frequency-comb (SPFC) supermodes
with deterministic phase control~\cite{Lee2018,Yoon2021}.
These supermodes constitute well-defined bosonic modes whose temporal
and spectral structure is inherited directly from a stabilized
optical frequency comb.
Because the same metrological reference defines both state generation
and displacement operations,
logical stabilizers, logical Pauli operators, and physical control knobs
share a common operational foundation.
This hardware-native correspondence distinguishes the present platform
from abstract phase-space constructions and from discrete time-bin or frequency-bin encodings~\cite{Marcikic2004,Xavier2025,Lu2023,Kues2019},
which store logical information in labeled mode subspaces rather
than in a stabilizer lattice embedded in continuous time--frequency
phase space.
 
A realistic assessment of the associated error structure is essential.
In the TF domain, dominant imperfections such as timing jitter,
phase noise, finite bandwidth, and dispersion
naturally act as approximately Gaussian displacement errors in the
logical phase space~\cite{Fabre2022,Ren2014,Descamps2023b}.
Within experimentally accessible regimes,
these shifts remain small compared to the stabilizer lattice spacing,
yielding intrinsic correctability in the standard GKP
shift-noise picture~\cite{Fukui2018,Noh2020a,Grimsmo2021,Seshadreesan2022}.
We quantify this displacement-noise model directly in terms of measurable
laboratory parameters and delineate the regime of passive protection.

Beyond intrinsic resilience, the time--frequency displacement algebra
that governs noise processes also underlies all coherent control in the
platform.
Phase and delay manipulations implement calibrated TF displacements,
which realize logical Pauli generators and continuous logical rotations
without additional hardware overhead.
Building on the modal time--frequency GKP framework
\cite{Descamps2024a},
we further identify a concrete pathway toward active syndrome extraction
and deterministic displacement recovery using ancillary grid states \cite{Fukui2018,Larsen2021a},
interferometric coupling \cite{Gottesman2001,Larsen2021a}, and time--frequency-resolved detection \cite{Fabre2022,Descamps2024a}.
These primitives establish the operator-level and hardware-level
ingredients required for integration into erasure-aware and
fusion-based fault-tolerant photonic architectures.

Finally, the TF platform is inherently compatible with scalability.
Frequency-comb structures and shared metrological references enable
parallel encoding of multiple logical qubits across distinct TF modes,
providing a natural route toward multiplexed photonic architectures
\cite{Kues2019,Lu2023,Wang2018}.
While we do not experimentally demonstrate repeated error-correction cycles
or establish a threshold for fault tolerance,
the present work realizes the logical encoding, stabilizer structure,
noise model, control operations, and syndrome-extraction primitives
necessary for a hardware-native time--frequency GKP layer
toward fault-tolerant photonic operation.

The remainder of this paper is organized as follows.
Section~\ref{sec:TFphase} introduces the time--frequency phase space and the
TF displacement operators underlying the GKP mapping.
Sections~\ref{sec:logical_qubit}--\ref{sec:scalability} develop the
hardware-native TF--GKP encoding, including the logical subspace and Pauli
operators, metrologically enforced stabilizers, finite-energy grid-state models,
a quantitative displacement-noise framework, deterministic logical operations,
and multiplexed scalability.
Section~\ref{sec:discussion} discusses physical error channels and architectural
implications for time--frequency photonics.
Section~\ref{sec:ft_extension} then outlines a concrete pathway toward active
time--frequency GKP syndrome extraction and deterministic recovery, positioning
the platform toward fault-tolerant photonic operation.
Section~\ref{sec:conclusion} concludes, and the Appendices provide detailed
derivations and feasibility estimates.

\begin{figure*}[t]
\centering
\includegraphics[width=\linewidth]{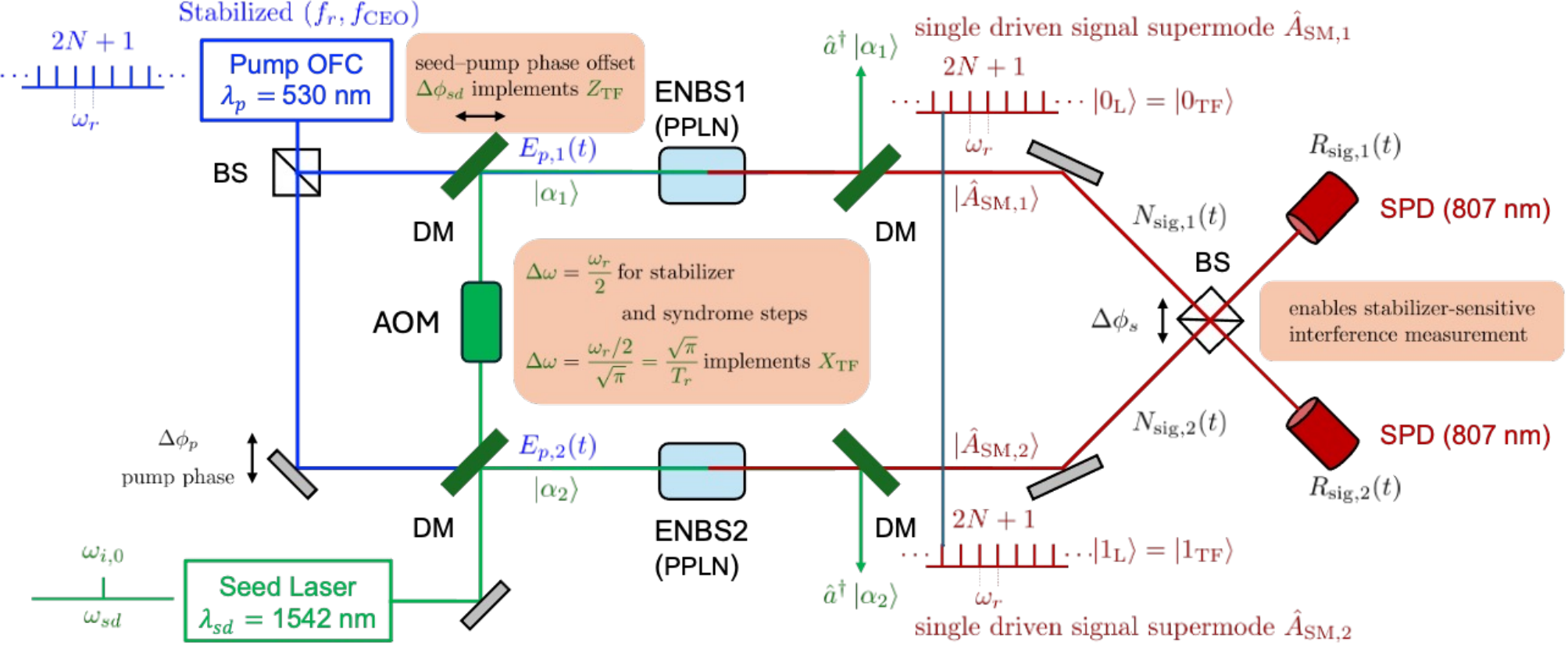}
\caption{\textbf{Hardware-native implementation of a time--frequency (TF) GKP logical qubit using coherently seeded entangled nonlinear biphoton sources.}
A stabilized optical frequency comb (OFC) with locked repetition rate $f_r$ and carrier-envelope offset frequency $f_{\mathrm{CEO}}$ pumps two coherently seeded entangled nonlinear biphoton sources (ENBS1/ENBS2) implemented in periodically poled lithium niobate (PPLN) crystals, thereby defining a metrologically anchored TF lattice with spacing $(T_r,\omega_r)$.
A phase-locked idler seed laser selects pump-matched driven signal supermodes $\ket{\hat A_{\mathrm{SM},1}}$ and $\ket{\hat A_{\mathrm{SM},2}}$ (red), which constitute the physical bosonic modes encoding the logical states $\ket{0_L}=\ket{0_{\mathrm{TF}}}$ and $\ket{1_L}=\ket{1_{\mathrm{TF}}}$.
Control parameters map directly to TF displacement operators: the pump--seed phase offset $\Delta\phi_{sd}$ implements $\bar{Z}$-type temporal translations, while an acousto-optic modulator (AOM) applies calibrated spectral displacements $\Delta\omega=\omega_r/2$ (equivalently $\sqrt{\pi}/T_r$ under the GKP mapping) corresponding to $\bar{X}$-type logical translations and stabilizer or syndrome increments.
A TF interferometer with adjustable signal-phase shift $\Delta\phi_s$ enables stabilizer-sensitive interference measurements of $\langle S_\tau\rangle$ and $\langle S_\Omega\rangle$.
The port-resolved signal outputs $N_{\mathrm{sig},1}(t)$ and $N_{\mathrm{sig},2}(t)$ are detected using single-photon detectors (SPD) at the signal wavelength (807\,nm).
}\label{fig:fig1}
\end{figure*}

\section{Time--Frequency Phase Space and GKP Operators}
\label{sec:TFphase}

The TF degrees of freedom of a single photon admit an
operational continuous-variable phase-space description when the photon
occupies a single, well-defined spatiotemporal mode.
This chronocyclic phase-space description has a long history in ultrafast and
quantum optics, where it underpins time--energy uncertainty relations,
temporal-mode decompositions, and frequency-comb physics
\cite{Ren2014,Descamps2023b,Raymer1994,Brecht2015, Fabre2020}.
Here, we exploit this structure to construct a hardware-native realization of
Gottesman--Kitaev--Preskill (GKP) encoding directly in the TF domain.

\subsection{Canonical time--frequency operators}

We describe the signal photon in terms of a single, well-defined temporal mode
envelope, for which the arrival-time operator $\hat{t}$ and angular-frequency
operator $\hat{\omega}$ satisfy the canonical commutation relation
\begin{equation}
[\hat{t},\hat{\omega}] = i,
\label{eq:TF_commutation}
\end{equation}
in units where $\hbar=1$.
This relation is formally equivalent to the position--momentum commutator
$[\hat{q},\hat{p}]=i$ and is widely used in chronocyclic quantum optics when the
photon occupies a single spatiotemporal mode
\cite{Ren2014,Brecht2015,Fabre2020}.
This effective canonical structure arises after restricting to a single
orthogonal temporal supermode and is valid within the ultralow-gain,
single-photon manifold considered here.

Physical displacements in TF phase space are generated by the Weyl operator
\begin{equation}
\hat{D}_{\mathrm{TF}}(\Delta t,\Delta\omega)
=
\exp\!\left[i\bigl(\Delta\omega\,\hat{t}
-
\Delta t\,\hat{\omega}\bigr)\right],
\label{eq:TF_displacement}
\end{equation}
which implements a temporal delay $\Delta t$ and a frequency translation
$\Delta\omega$.
Such displacements are experimentally realized using optical delay lines,
electro-optic or acousto-optic modulation, and phase-controlled interference,
as summarized schematically in Fig.~\ref{fig:fig1}.

\subsection{Dimensionless canonical variables}

To connect the physical TF phase space to a metrologically stabilized lattice,
we introduce dimensionless canonical operators
\begin{equation}
\hat{\tau} = \frac{\hat{t}}{T_r},
\qquad
\hat{\Omega} = T_r \hat{\omega},
\label{eq:dimensionless_TF}
\end{equation}
where $T_r=1/f_r$ is the repetition period of a stabilized optical frequency
comb.
These operators satisfy $[\hat{\tau},\hat{\Omega}]=i$ and define a dimensionless
phase space for a single effective signal temporal mode.
The explicit reduction to this driven signal supermode is derived in
Appendix~\ref{app:supermode}.

The corresponding dimensionless displacement operator is
\begin{equation}
\hat{D}(\tau,\Omega)
=
\exp\!\left[i(\Omega\hat{\tau}-\tau\hat{\Omega})\right],
\label{eq:dimensionless_displacement}
\end{equation}
which generates physical shifts
$\Delta t = \tau T_r$ and
$\Delta\omega = \Omega/T_r$.
Because both the repetition rate $f_r$ and the carrier-envelope offset
frequency $f_{\mathrm{CEO}}$ can be stabilized with sub-Hz precision
\cite{Udem2002,Cundiff2003,Hall2006,Haensch2006},
the TF phase space is not merely formal but metrologically anchored,
with lattice spacings defined directly by experimentally stabilized
comb parameters.

\subsection{GKP stabilizers in the TF phase space}

Within this dimensionless TF phase space, the standard square-lattice GKP code
is defined by the commuting stabilizers
\begin{equation}
\hat{S}_{\tau}
=
\hat{D}(0,2\sqrt{\pi}),
\qquad
\hat{S}_{\Omega}
=
\hat{D}(2\sqrt{\pi},0),
\label{eq:GKP_stabilizers}
\end{equation}
which fix a discrete lattice in $(\tau,\Omega)$ space
\cite{Gottesman2001,Descamps2024a,Menicucci2014}.
The logical unit cell has area $2\pi$, and logical Pauli operators correspond to
half-lattice translations.

\begin{figure}[t]
\centering
\includegraphics[width=\linewidth]{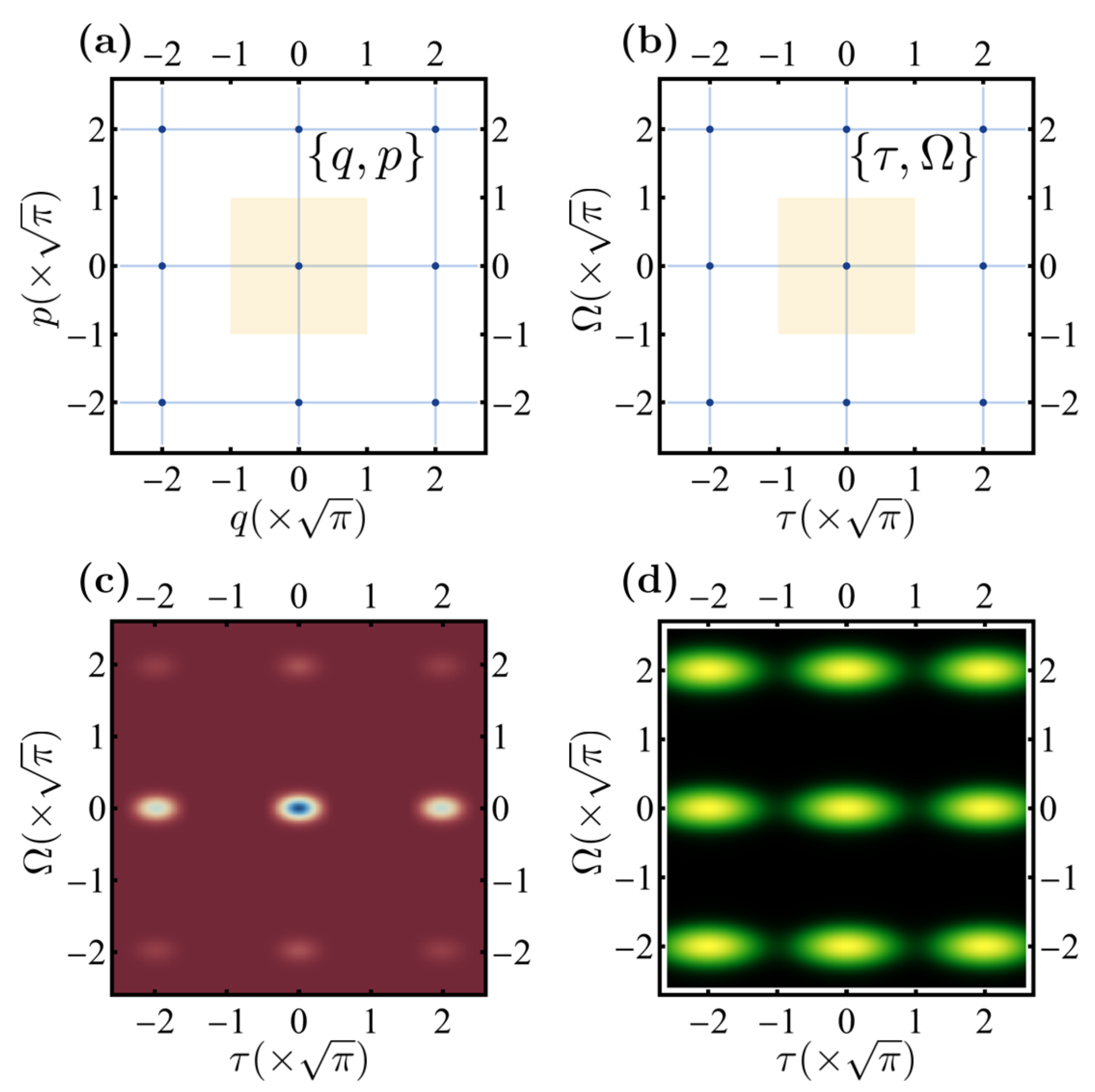}
\caption{\textbf{From canonical GKP codes to TF--GKP encoding and finite-energy grid-state models.}
(\textbf{a}) Canonical square-lattice GKP code in the $(q,p)$ phase space, shown in units of $\sqrt{\pi}$ with lattice spacing $2\sqrt{\pi}$ and the logical unit cell highlighted.
(\textbf{b}) Corresponding TF--GKP lattice in the dimensionless canonical variables $(\tau,\Omega)$ defined in Sec.~\ref{sec:TFphase}, illustrating the isomorphic operator mapping $(\hat q,\hat p)\leftrightarrow(\hat\tau,\hat\Omega)$.
(\textbf{c}) Representative Wigner-function illustration of a finite-energy TF grid state, showing localized lattice peaks and interference structure determined by finite effective squeezing.
(\textbf{d}) Comb-of-Gaussians approximation with broadened lattice peaks characterized by widths $(\sigma_\tau,\sigma_\Omega)$, used to quantify intrinsic displacement tolerance and logical failure probability in Sec.~\ref{sec:errors} and Fig.~\ref{fig:fig4}(a).
} \label{fig:fig2}
\end{figure}

Mapping Eq.~\eqref{eq:GKP_stabilizers} back to physical units yields stabilizer
periods
\begin{equation}
\Delta t_{\mathrm{GKP}} = 2\sqrt{\pi}\,T_r,
\qquad
\Delta\omega_{\mathrm{GKP}} = \frac{2\sqrt{\pi}}{T_r},
\label{eq:GKP_physical_periods}
\end{equation}
demonstrating that the GKP lattice spacing is set directly by the comb repetition
rate.
This direct correspondence between stabilized comb parameters and GKP
stabilizer periods is illustrated in Fig.~\ref{fig:fig2}(a,b),
highlighting that the logical lattice is defined by experimentally
calibrated displacements rather than abstract quadrature scalings.

\subsection{Finite-energy TF grid states}

Ideal GKP code states are unphysical, requiring infinite energy and perfect
delta-function localization.
Physical realizations therefore employ approximate grid states with finite
width in both conjugate quadratures
\cite{Gottesman2001,Fukui2018,Menicucci2014}.
In the TF setting, such states correspond to a lattice of temporally and
spectrally localized peaks with Gaussian envelopes, characterized by effective
noise widths $(\sigma_{\tau},\sigma_{\Omega})$.

A convenient representation is the comb-of-Gaussians model, in which the Wigner
function exhibits a periodic lattice with finite peak widths and interference
fringes
\cite{Weedbrook2012,Albert2018}.
Representative TF grid-state structures are shown in Fig.~\ref{fig:fig2}(c,d).
The quantitative connection between $(\sigma_{\tau},\sigma_{\Omega})$ and logical
failure probability is analyzed in Sec.~\ref{sec:errors}, with full derivations
provided in Appendices~C and~D.

These finite-energy parameters $(\sigma_\tau,\sigma_\Omega)$ not only
quantify passive displacement tolerance but also determine the syndrome
resolution required for active correction protocols discussed in
Sec.~\ref{sec:ft_extension}.

\subsection{Scope and forward connection}

This section establishes the operator-level foundation of TF--GKP encoding:
(i) time and frequency define an effective canonical phase space for a
single temporal supermode,
(ii) stabilized frequency-comb parameters provide dimensionless TF
coordinates anchored to laboratory references, and
(iii) the square-lattice GKP stabilizers map directly onto calibrated
time and frequency displacements.
The logical lattice is therefore defined natively at the hardware level,
providing the structural layer upon which passive protection and
active syndrome-extraction pathways are subsequently built.

In the following sections, we use this TF phase-space framework to construct
finite-energy TF--GKP code states, analyze their noise sensitivity, and connect
logical operations and measurements explicitly to experimentally accessible
TF control primitives.

\section{Time--Frequency Bosonic Logical Qubit}
\label{sec:logical_qubit}

Having established the time--frequency (TF) phase space and its canonical
displacement operators in Sec.~\ref{sec:TFphase}, we now define a
\emph{bosonic logical qubit} encoded within this space.
Our goal is to specify the protected logical subspace, identify a logical basis,
and define the logical Pauli operators, following the Gottesman--Kitaev--Preskill
(GKP) framework while maintaining a direct connection to experimentally
accessible TF displacements.

\subsection{Logical subspace defined by TF stabilizers}

A bosonic logical qubit is defined by restricting the infinite-dimensional
Hilbert space of the TF degree of freedom to a discrete lattice stabilized by
commuting displacement operators.
Specifically, we consider the stabilizers
$\hat{S}_{\tau} = \hat{D}(0,2\sqrt{\pi})$ and
$\hat{S}_{\Omega} = \hat{D}(2\sqrt{\pi},0)$,
as introduced in Sec.~\ref{sec:TFphase}.
The logical code space is the simultaneous $+1$ eigenspace of
$\hat{S}_{\tau}$ and $\hat{S}_{\Omega}$, and corresponds to a square lattice in
the $(\tau,\Omega)$ phase space with unit-cell area $2\pi$
\cite{Gottesman2001,Descamps2024a,Menicucci2014}.

\subsection{Logical basis states}

The logical qubit basis $\{\ket{0_L},\ket{1_L}\}$ is defined by two inequivalent
offsets of the TF grid relative to the stabilizer lattice.
Concretely, the logical states may be chosen as eigenstates of the stabilizers
with distinct eigenvalues under half-lattice translations,
\begin{align}
\ket{0_L} &: \text{grid centered at } (\tau,\Omega) = (0,0), \\
\ket{1_L} &: \text{grid centered at } (\tau,\Omega) = (\sqrt{\pi},0),
\end{align}
up to an overall phase convention.
This choice mirrors the canonical position-space realization of GKP codewords,
with the TF variables playing the role of conjugate quadratures.

In practice, physical realizations correspond to approximate grid states with
finite peak widths and Gaussian envelopes, as illustrated in
Fig.~\ref{fig:fig2}(c,d).
These finite-energy states retain the essential logical structure provided that
their effective noise widths remain well below the stabilizer half-spacing
\cite{Fukui2018,Grimsmo2021}.
A quantitative treatment of finite-energy effects is given in Appendices~C and~D.

\subsection{Logical Pauli operators}

Logical operators acting within the encoded subspace correspond to
half-lattice TF displacements that commute with the stabilizers
but act nontrivially on the logical basis.
The logical Pauli operators are identified as
\begin{equation}
\bar{Z} = \hat{D}(\sqrt{\pi},0),
\qquad
\bar{X} = \hat{D}(0,\sqrt{\pi}),
\label{eq:logical_Paulis}
\end{equation}
up to overall phases.
Within the code space, $\bar{X}$ exchanges the logical basis states
$\ket{0_L}\leftrightarrow\ket{1_L}$, while $\bar{Z}$ imparts a relative phase
between them, and both commute with the stabilizers.

\begin{figure*}[t]
\centering
\includegraphics[width=\linewidth]{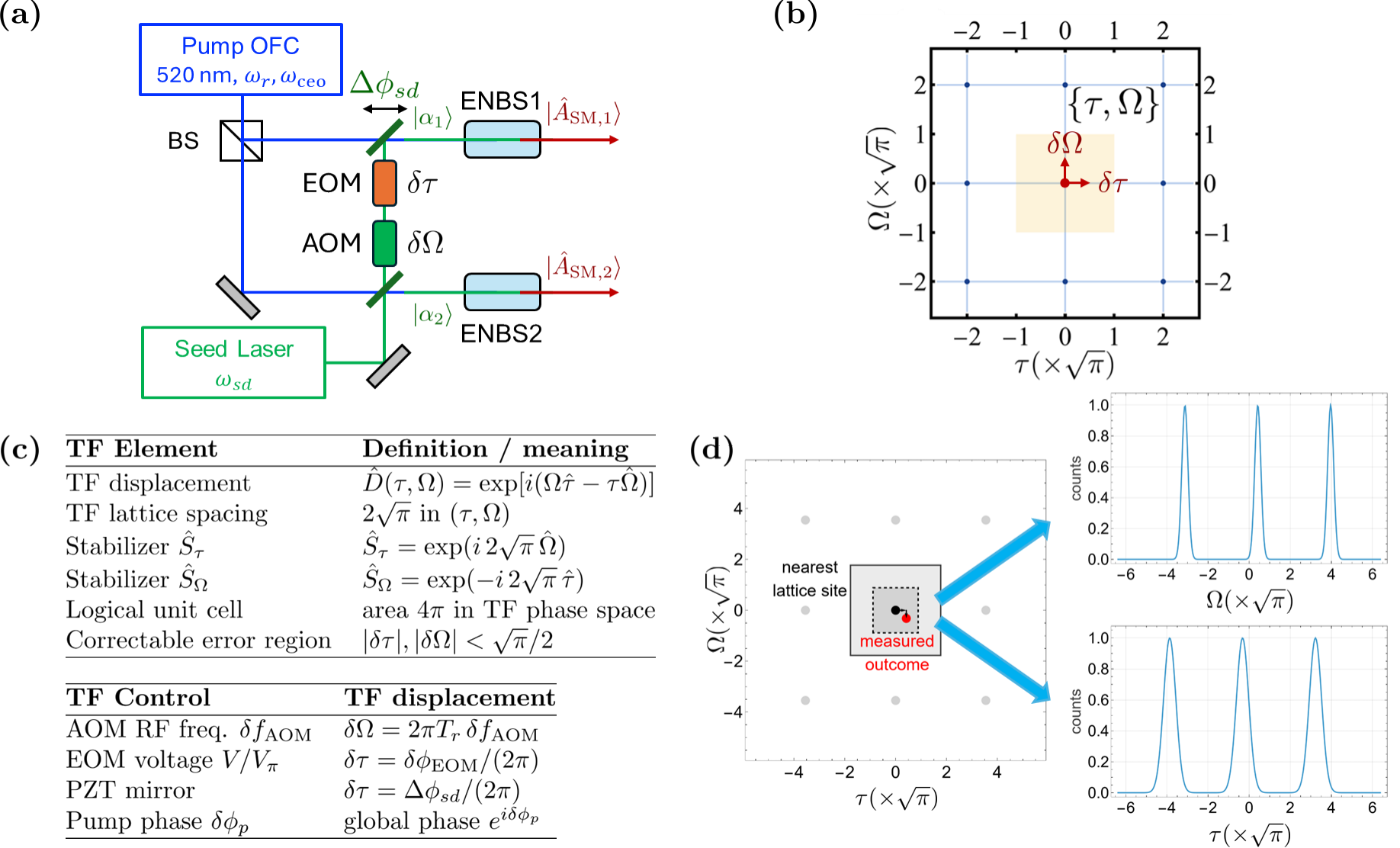}
\caption{\textbf{Operational mapping between laboratory controls and TF--GKP displacements and syndrome inference.}
(\textbf{a}) Control architecture: a stabilized pump optical frequency comb (OFC) and phase-locked seed laser drive two ENBS units, while an electro-optic modulator (EOM) and an acousto-optic modulator (AOM) implement programmable time--frequency (TF) displacement operations.
(\textbf{b}) Displacement picture in the TF phase space $(\tau,\Omega)$: calibrated controls generate small logical or noise-induced displacements $(\delta\tau,\delta\Omega)$ within the shaded logical unit cell defined by the stabilizer lattice.
(\textbf{c}) Summary of TF--GKP code elements (displacement operators, lattice spacing $2\sqrt{\pi}$, stabilizers, logical unit cell, correctable region) together with calibrated control-to-displacement mappings (AOM RF frequency $\rightarrow \delta\Omega$, EOM voltage $\rightarrow \delta\tau$, PZT/pump-phase offsets $\rightarrow$ phase translations).
(\textbf{d}) Conceptual syndrome interpretation: a measured TF outcome is associated with the nearest lattice site within a stabilizer cell, and time- and frequency-resolved detection provide projections onto $\tau$ and $\Omega$ that supply the geometric information required for displacement decoding.
} \label{fig:fig3}
\end{figure*}

\subsection{Logical Bloch sphere and continuous operations}

Beyond discrete Pauli operations, the TF--GKP encoding naturally supports
continuous logical rotations.
Small displacements within a stabilizer unit cell correspond to correctable shifts that move the state within the same logical equivalence class, enabling continuous logical control within the finite-energy code space.
This feature is a general property of bosonic encodings and underlies their
utility for analog control and continuous-variable quantum information
processing \cite{Weedbrook2012,Albert2018}.

In the TF platform, such continuous operations are realized by smoothly tuning
temporal delays or frequency offsets.
As long as these displacements remain within the correctable region defined by
the stabilizer lattice, logical coherence is preserved.
This enables deterministic logical control at the encoded level
without requiring measurement-induced postselection,
while remaining compatible with active correction protocols
discussed in Sec.~\ref{sec:ft_extension}.

\subsection{Relation to lower-dimensional encodings}

The TF--GKP logical qubit generalizes lower-dimensional logical constructions
previously demonstrated on the ENBS/SPFC platform.
A previously explored two-mode time–frequency qubit realization \cite{Yoon2025b} can be viewed, in retrospect, as the lowest-dimensional limit of the present TF–GKP framework, in which only two TF alternatives are accessed without invoking a full lattice structure.
The present encoding extends this approach to a genuinely bosonic code by
exploiting the periodic TF structure inherited from the stabilized frequency
comb.

This hierarchy clarifies the conceptual role of the TF--GKP encoding:
rather than introducing a qualitatively new physical resource, it elevates
existing phase-engineered TF interference into a lattice-protected bosonic
logical subspace.

\subsection{Logical-qubit architecture and scaling.}
In the present implementation, a single TF--GKP logical qubit is encoded using
two coherently driven ENBS units whose signal supermodes are interferometrically
combined under a shared metrological reference.
This bosonic generalization preserves the continuous-variable stabilizer
structure required by modal GKP fault-tolerance schemes.
This construction scales naturally: an array of $2N$ ENBS units, grouped into
$N$ independently addressed pairs and phase referenced to the same optical
frequency comb, realizes $N$ parallel TF--GKP logical qubits.

\subsection{Summary and forward connection}

In this section we have defined a bosonic logical qubit encoded in the TF phase
space of a single photon, specified its stabilizer-defined subspace, logical
basis states, and Pauli operators, and connected each element directly to
physical TF displacements.
The logical structure mirrors the canonical GKP code while remaining fully
hardware-native.

In the next section, we show how these stabilizers are not merely algebraic
constructs but are physically enforced by metrological reference anchoring,
providing continuous logical-state pinning and establishing the displacement
noise model that underlies both intrinsic protection and the active
fault-tolerance pathway developed in Sec.~\ref{sec:ft_extension}.

\section{Stabilizer Structure from Metrological Reference Anchoring}
\label{sec:stabilizers}

A defining feature of the TF bosonic logical qubit introduced
in Sec.~\ref{sec:logical_qubit} is that its stabilizer structure is not imposed
abstractly, but instead arises from experimentally enforced phase coherence.
In this section, we show how an external metrological reference anchors the TF
phase space and thereby defines a set of commuting displacement stabilizers that
pin the logical subspace continuously in time.

\subsection{Stabilizers in abstract GKP codes}

In the canonical GKP framework, stabilizers are defined algebraically as
commuting displacement operators that generate a lattice in phase space
\cite{Gottesman2001}.
These operators define the logical code space but do not, by themselves, specify
a physical mechanism by which stabilizer eigenvalues are enforced or maintained.
In most theoretical treatments, stabilizers are assumed to be exact symmetries
of the encoded states, while in experimental implementations they are enforced
only approximately, often through measurement and feedback
\cite{Albert2018,Terhal2020,Grimsmo2021}.

This separation between algebraic definition and physical enforcement is a
well-known challenge in bosonic encodings. It motivates active error-correction protocols in which stabilizer violations
are detected and corrected dynamically.
By contrast, the approach taken here embeds stabilizer enforcement directly into
the hardware through phase referencing, providing a complementary route to
logical protection.

\subsection{Phase-referenced TF lattice as a physical stabilizer}

As established in Sec.~\ref{sec:TFphase}, the TF phase space of the driven ENBS
supermode is rigidly anchored to stabilized optical-frequency-comb parameters,
specifically the repetition period $T_r$ and angular spacing $\omega_r$.
Because these quantities are phase locked to an external reference with
sub-femtosecond and sub-Hz precision \cite{Udem2002,Cundiff2003,Hall2006},
TF displacements differing by integer multiples of $(T_r,\omega_r)$ acquire
identical physical phases.

This anchoring ensures that TF displacements separated by integer
multiples of the stabilizer periods $(2\sqrt{\pi},0)$ and $(0,2\sqrt{\pi})$
acquire well-defined and reproducible phases relative to the shared
metrological reference.
Within the encoded subspace, the corresponding displacement operators
$\hat{S}_{\tau}=\hat{D}(0,2\sqrt{\pi})$ and
$\hat{S}_{\Omega}=\hat{D}(2\sqrt{\pi},0)$
act as stabilizers in the sense that states related by their action
belong to the same logical equivalence class.
The metrological anchoring therefore promotes the abstract GKP
stabilizers to physically calibrated displacement symmetries.
They therefore act as \emph{physical stabilizers} of the logical subspace, rather
than merely algebraic ones.
Within the encoded logical subspace, phase-coherent measurements that
respect the shared reference do not distinguish states differing by
stabilizer displacements.

Operationally, this means that the stabilizer conditions are enforced
continuously by the experimental configuration itself, without requiring
explicit syndrome extraction.
The physical origin of this enforcement is illustrated schematically in
Fig.~\ref{fig:fig1} and quantified in Appendices~A and~E.

\subsection{Continuous pinning of the logical subspace}

Because all driven supermodes share a common phase reference, slow
drifts in optical path length, pump phase, or seed phase manifest as
global TF displacements that translate the entire stabilizer lattice. Such global shifts commute with the stabilizers and therefore preserve logical equivalence classes.
Only relative displacements within a stabilizer unit cell modify the
encoded logical information.

This continuous pinning of the logical subspace distinguishes the TF platform
from implementations in which stabilizers are enforced intermittently through
measurement.
Here, the logical equivalence classes are defined at the hardware level:
all TF states within a stabilizer unit cell encode the same logical information.
Small perturbations therefore move the physical state within the same logical
class, rather than ejecting it from the code space.

\subsection{Relation to displacement errors and correctability}

The stabilizer structure enforced by metrological anchoring defines the natural
error model of the encoding.
Timing jitter, phase noise, and frequency fluctuations act as stochastic TF
displacements.
Provided these displacements remain smaller than half the stabilizer spacing,
they do not change the stabilizer eigenvalues and are therefore correctable at
the encoding level.

This formulation places the TF encoding directly within the standard
Gaussian shift-noise framework used in bosonic fault-tolerance analyses
\cite{Fukui2018,Noh2020a,Noh2020b,Larsen2021a}.
In the TF platform, the noise model is not assumed but derived directly from the
dominant physical imperfections of the hardware.
A quantitative mapping between noise widths and logical failure probability is
presented in Fig.~\ref{fig:fig4}(a) and derived explicitly in Appendices~C and~D.

\begin{figure*}[t]
\centering
\includegraphics[width=0.9\linewidth]{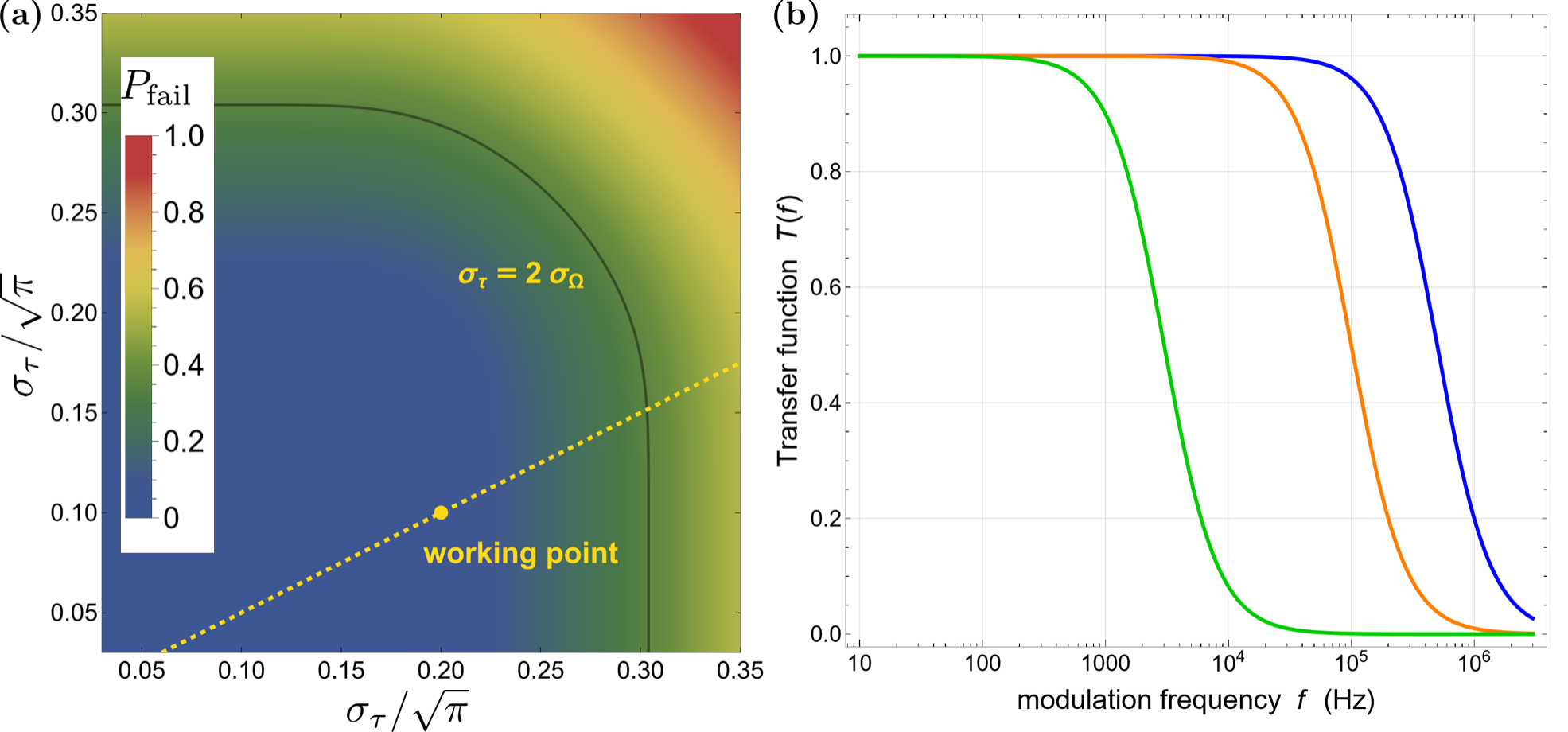}
\caption{\textbf{Quantitative robustness and control feasibility for TF--GKP encoding.}
(\textbf{a}) Logical failure probability $P_{\mathrm{fail}}$ under nearest-lattice decoding of a finite-energy TF grid state subject to Gaussian displacement noise, plotted versus normalized noise widths $\sigma_{\tau}/\sqrt{\pi}$ and $\sigma_{\Omega}/\sqrt{\pi}$.
The map exhibits a broad low-error region (blue), with contours of constant $P_{\mathrm{fail}}$ indicating the intrinsic correctability landscape for simultaneous timing and frequency fluctuations.
The dashed line $\sigma_{\tau}=2\sigma_{\Omega}$ marks a representative anisotropic-noise (or anisotropic-squeezing) condition naturally compatible with the ENBS/SPFC platform, and the highlighted point indicates a realistic operating regime within the high-fidelity region.
(\textbf{b}) Representative transfer functions $T(f)$ of experimentally available actuation channels (piezoelectric transducer (PZT), acousto-optic modulator (AOM), electro-optic modulator (EOM)), showing that the control bandwidth required to implement TF displacements for logical operations and stabilizer/syndrome steps is compatible with available phase and frequency modulation technology.
} \label{fig:fig4}
\end{figure*}

\subsection{Distinction from active stabilization schemes}

It is important to distinguish metrological stabilizer anchoring from active
error correction.
No measurements of stabilizer eigenvalues are performed in the present
implementation, nor is any feedback applied to correct detected errors.
Instead, the stabilizers define equivalence classes of physical states that are
continuously enforced by phase coherence.

This passive enforcement does not by itself establish a fault-tolerance
threshold.
However, it realizes the hardware-level stabilizer structure assumed in modal
GKP fault-tolerant architectures and provides a calibrated displacement
algebra upon which active syndrome extraction can be layered.
In this sense, the TF encoding realizes the ``hardware layer'' of a GKP code,
while leaving open multiple pathways toward full fault tolerance.

\subsection{Summary and forward connection}

In summary, the stabilizers defining the TF bosonic logical qubit are
physically calibrated through metrological reference anchoring.
This mechanism pins the TF lattice, suppresses uncontrolled relative
phase drift, and ensures that dominant physical noise processes map
onto small displacement errors within a well-defined stabilizer cell.
While this passive enforcement does not by itself establish a
fault-tolerance threshold, it provides the hardware-native stabilizer
infrastructure required for the active syndrome extraction and
recovery strategies developed in Sec.~\ref{sec:ft_extension}.

\section{Error Structure and Intrinsic Correctability}
\label{sec:errors}

A central motivation for bosonic encodings is that physically relevant noise
processes can be mapped onto structured errors in phase space, where small
perturbations correspond to correctable displacements.
In this section, we analyze the dominant error channels affecting the
TF bosonic logical qubit and show that, within an experimentally
accessible regime, these processes correspond to intrinsically correctable
displacement errors protected by the stabilizer structure established in
Sec.~\ref{sec:stabilizers}.

\subsection{Physical noise channels in the TF platform}

In the ENBS/SPFC architecture, the dominant imperfections arise from three
sources: (i) timing jitter, (ii) phase and frequency noise, and (iii) photon
loss.
Timing jitter produces fluctuations in the arrival time of the signal photon,
while phase and frequency noise induce fluctuations in the conjugate spectral
degree of freedom.
Both effects act as stochastic displacements in the TF phase space and are
naturally described by the displacement operators introduced in
Sec.~\ref{sec:TFphase}.

Photon loss constitutes a qualitatively different error channel.
Loss removes the excitation from the logical subspace entirely and cannot be
corrected within the present encoding.
However, loss events can be identified through standard photonic detection and
treated as erasures at the architectural level.
Because the focus of this work is on intrinsic protection against small
continuous errors, we concentrate primarily on displacement-type noise in the
following analysis.

\subsection{Displacement-noise model}

Small timing and frequency fluctuations induce random displacements
$\hat{D}(\delta\tau,\delta\Omega)$ in the dimensionless TF phase space.
In realistic operating conditions, these fluctuations are well modeled by a
Gaussian distribution with variances $(\sigma_\tau^2,\sigma_\Omega^2)$,
\begin{equation}
P(\delta\tau,\delta\Omega)
=
\frac{1}{2\pi\sigma_\tau\sigma_\Omega}
\exp\!\left[
-\frac{\delta\tau^2}{2\sigma_\tau^2}
-\frac{\delta\Omega^2}{2\sigma_\Omega^2}
\right].
\end{equation}
This noise model coincides with the Gaussian-shift channels commonly assumed in
bosonic fault-tolerance analyses
\cite{Fukui2018,Noh2020a,Grimsmo2021,Noh2020b,Larsen2021a}.

Importantly, in the TF platform this model is not postulated but derived directly
from the dominant physical noise sources.
Timing jitter of the pump comb contributes primarily to $\sigma_\tau$, while
seed-phase and frequency noise contribute to $\sigma_\Omega$.
Both parameters are experimentally tunable through reference stabilization and
dispersion management, as discussed in Sec.~\ref{sec:stabilizers} and quantified
in Appendices~C and~D.

\subsection{Intrinsic correctability from stabilizer structure}

The stabilizer lattice enforced by metrological reference anchoring partitions
the TF phase space into equivalent logical cells.
Displacements smaller than half the stabilizer spacing remain within a
single logical cell and are correctable under nearest-lattice decoding.
In the ideal infinite-energy limit, such shifts leave the logical information
invariant; for finite-energy states, the logical error probability decreases
rapidly as the displacement width becomes small compared to the half-cell size.

This intrinsic correctability is a direct consequence of the stabilizer
structure and does not require active syndrome extraction in the present
demonstration.
Instead, the logical information is encoded in equivalence classes of physical
states related by stabilizer displacements.
As long as the noise-induced displacements remain within the correctable region,
the logical state is preserved automatically.

This mechanism mirrors the passive protection assumed in theoretical GKP codes,
but here it is realized physically through reference anchoring.
The correctability condition is therefore expressed as a quantitative constraint
on the noise widths $(\sigma_\tau,\sigma_\Omega)$ relative to the stabilizer
spacing.

\subsection{Noise scale and logical failure probability}

Finite-energy TF grid states possess broadened peaks, and sufficiently large
displacements can cause overlap between neighboring lattice sites, resulting in
logical errors.
The probability of such events defines the logical failure probability
$P_{\mathrm{fail}}$.

In Fig.~\ref{fig:fig4}(a), we plot $P_{\mathrm{fail}}$ as a function of the
normalized noise widths $\sigma_\tau/\sqrt{\pi}$ and
$\sigma_\Omega/\sqrt{\pi}$, using a finite-energy TF grid-state model detailed in
Appendices~C and~D.
The resulting landscape exhibits a broad low-error region in which simultaneous
timing and frequency noise remain well below the logical-error threshold.

A key feature of the ENBS/SPFC platform is that the noise is naturally anisotropic:
temporal fluctuations are typically larger than spectral ones due to finite
pulse width and dispersion.
This anisotropy corresponds to elliptical noise contours in TF phase space and
is explicitly captured in the model.
As shown by the dashed line in Fig.~\ref{fig:fig4}(a), realistic operating points
lie well within the high-fidelity regime.

\subsection{Limits of intrinsic protection}

While the stabilizer structure provides passive protection against small
displacement errors, it does not address all noise processes.
Large displacements exceeding half the stabilizer spacing produce logical
errors, and photon loss removes the state from the logical subspace entirely.
Moreover, without repeated syndrome extraction, noise accumulation over long
times will eventually degrade the logical state, motivating the active-correction
pathway outlined in Sec.~\ref{sec:ft_extension}.

These limitations are intrinsic to any passive bosonic encoding.
The present work therefore does not claim a fault-tolerance threshold or full
quantum error correction.
Instead, it establishes that the dominant physical noise channels in the TF
platform map naturally onto correctable displacement errors, consistent with the shift-noise assumptions underlying bosonic
fault-tolerance analyses.

\subsection{Role within a fault-tolerant architecture}

The intrinsic correctability demonstrated here constitutes the
hardware-native displacement layer of a GKP-based architecture.
Active syndrome extraction, repeated correction cycles, or concatenation
with higher-level codes may be layered above this encoding to achieve
full fault tolerance \cite{Terhal2020,Seshadreesan2022}.
Because the TF implementation realizes both the stabilizer structure
and a calibrated Gaussian displacement noise model, it satisfies the
structural prerequisites assumed in modal GKP fault-tolerant schemes
and provides the physical substrate for the active correction pathway
outlined in Sec.~\ref{sec:ft_extension}.

In the next section, we show how deterministic phase and delay controls implement
logical operations within the protected subspace, completing the definition of
the TF bosonic logical qubit.

\section{Logical Operations via Deterministic Phase and Delay Control}
\label{sec:operations}

Having established the logical subspace (Sec.~\ref{sec:logical_qubit}), the
stabilizer structure enforced by metrological reference anchoring
(Sec.~\ref{sec:stabilizers}), and the intrinsic correctability of dominant noise
channels (Sec.~\ref{sec:errors}), we now describe how logical operations are
implemented within the TF bosonic encoding.
A defining feature of the present platform is that logical gates arise directly
from deterministic phase and delay controls that are already available at the
level of state generation and interference, without introducing additional
hardware overhead or measurement-induced feedback.

\subsection{Displacement-based logical operations}

As shown in Sec.~\ref{sec:logical_qubit}, logical Pauli operators correspond to
half-lattice displacements in TF phase space, with
$\bar{Z}=\hat{D}(\sqrt{\pi},0)$ and
$\bar{X}=\hat{D}(0,\sqrt{\pi})$,
up to overall phases.
These operators commute with the stabilizers while acting nontrivially within
the logical subspace, and therefore constitute valid logical operations.

In the ENBS/SPFC architecture, these displacements are implemented through
deterministic control of optical phase and frequency.
A calibrated shift of the relative pump--seed phase produces a controlled
temporal displacement of the signal envelope, implementing $\bar{Z}$.
Conversely, a controlled frequency translation of the idler seed or signal field
produces a spectral displacement, implementing $\bar{X}$ (Fig.~\ref{fig:fig1}).
Because these operations correspond to calibrated TF displacement
operators, their logical action is determined at the operator level.
In practice, finite pulse shapes and envelopes influence the effective
grid-state width and therefore the achieved logical fidelity, but do
not modify the underlying displacement algebra.
Concretely, the AOM-imposed frequency step implements $\delta\Omega$ and thus realizes $\bar{X}$-type translations, while calibrated phase/delay control implements $\delta\tau$ and thus $\bar{Z}$-type translations (Fig.~\ref{fig:fig3}).

\subsection{Continuous logical rotations within a stabilizer cell}

Beyond discrete Pauli operations, the TF encoding naturally supports continuous
logical rotations.
Small displacements $(\delta\tau,\delta\Omega)$ within a stabilizer
unit cell generate continuous transformations of the encoded state.
While such displacements do not strictly commute with the stabilizers,
their effect remains within the correctable region and can be interpreted,
under nearest-lattice decoding, as continuous logical rotations.
Operationally, such transformations are implemented by smoothly tuning the
applied phase or delay over a fraction of the stabilizer period.

This capability is particularly natural in the TF platform, where phase and
delay can be adjusted continuously and reversibly using electro-optic or
acousto-optic modulation.
Within the intrinsic correctability regime identified in
Sec.~\ref{sec:errors}, these operations do not compromise logical protection, as
the resulting state remains confined to the same stabilizer equivalence class.
As a result, analog logical rotations can be performed without invoking
composite pulse sequences or digital gate synthesis.

A crucial requirement for any logical gate is that it preserve the stabilizer
structure defining the code space.
In the present implementation, both stabilizers and logical gates are generated
by TF displacement operators, and their commutation relations are therefore fixed
by the underlying Weyl--Heisenberg algebra
\cite{Weedbrook2012,Albert2018,Gottesman2001}.
Displacements corresponding to logical Pauli operators commute with the
stabilizers by construction, while small calibrated displacements within
a stabilizer cell remain correctable under the decoding procedure.

This hardware-level compatibility ensures that logical manipulation
does not introduce a qualitatively distinct error channel beyond the
underlying displacement noise already analyzed in
Sec.~\ref{sec:errors}.
Gate-induced imperfections therefore enter the same Gaussian
shift-noise model that governs intrinsic correctability.
In particular, gate-induced errors are not a separate category, but are absorbed
into the same displacement-noise model analyzed in Sec.~\ref{sec:errors}.
Logical gate fidelity is therefore governed by the same experimentally
controllable parameters that determine intrinsic correctability.

We emphasize that such continuous displacements are best viewed as calibrated
phase-space translations within a finite-energy code manifold; their logical
interpretation depends on the chosen decoding and on maintaining operation
within the correctable region (Sec.~\ref{sec:errors}).

\subsection{Operational mapping and experimental control}

The mapping between laboratory control parameters and logical TF displacements
is summarized in Fig.~\ref{fig:fig3}.
Phase modulation of the pump or signal path implements temporal displacements
$\delta\tau$, while acousto-optic or electro-optic modulation implements spectral
displacements $\delta\Omega$.
These controls are deterministic, phase referenced, and compatible with
MHz-scale bandwidths, as quantified in Fig.~\ref{fig:fig4}(b).

From an operational perspective, logical gates are therefore executed by
programming calibrated phase and frequency shifts, rather than by synthesizing
abstract unitary operations.
This direct mapping elevates standard optical control techniques to logical
operations within a protected bosonic subspace.

\subsection{Gate set and scope}

The logical operations discussed here constitute a native gate set determined by
the available TF displacement controls.
This set includes logical Pauli operations and continuous single-qubit
rotations, but does not by itself provide a universal logical gate set.
We emphasize that the goal of this work is not to demonstrate universal quantum
computation, but rather to establish a physically grounded logical qubit whose
operations are deterministic, hardware native, and compatible with intrinsic
error protection.

Extensions to larger gate sets, including multi-qubit interactions or
non-Gaussian logical operations, may be realized by coupling TF modes through
additional interferometric or nonlinear resources.
Such extensions are compatible with the present encoding but lie beyond the
scope of this work.

In the following section, we show how these features extend naturally to
multiplexed architectures, enabling the parallel realization of multiple TF
bosonic logical qubits within a shared metrological reference frame.

\section{Scalability and Multiplexed Logical Qubits}
\label{sec:scalability}

A central requirement for any quantum-information architecture is the ability
to scale beyond a single logical qubit while preserving control, coherence, and
logical protection.
The TF photonic platform developed here is intrinsically well
suited to this task, as it naturally supports multiplexing across distinct
spectral and temporal modes while maintaining a shared metrological reference.
In this section, we clarify the precise meaning of scalability in the present
context and delineate both its strengths and its scope.

\subsection{Frequency-comb multiplexing of logical qubits}

The TF--GKP encoding introduced in this work can be replicated across multiple,
spectrally distinct modes provided by an optical frequency comb.
Each selected comb mode defines an independent TF phase space with its own
stabilizer lattice and logical subspace, allowing multiple bosonic logical qubits
to be encoded in parallel.
Because the stabilizers and logical operators are defined through TF
displacements, their algebraic structure is identical for all modes.

Crucially, all logical qubits share the same stabilized comb parameters
$(f_r,f_{\mathrm{CEO}})$ and are therefore referenced to a common time--frequency
origin.
This shared reference ensures that the relative phase between different
logical qubits is globally defined with respect to the same stabilized
comb parameters and is therefore subject only to the residual instability
of the common metrological reference.
As a result, multiplexed logical qubits remain mutually phase coherent without
requiring independent stabilization or calibration for each mode.

In practical terms, each logical qubit is realized using a distinct entangled
nonlinear biphoton source (ENBS) pair operating at a selected frequency-comb
channel, while all ENBS units are pumped by the same stabilized optical frequency
comb and seeded by the same phase-locked idler laser.
This architecture does not require independent stabilization loops for
each logical qubit; all encoded modes inherit their phase reference
from the same stabilized comb.

Although the analysis here focuses on a single logical qubit encoded using two
ENBS units, the construction generalizes directly to $N$ logical qubits realized
by $2N$ ENBS units, with each qubit defined by a phase-referenced ENBS pair. 

\subsection{Parallel logical operations}

Logical operations based on phase and delay control act locally within each TF
mode and therefore extend naturally to multiplexed architectures.
Independent modulation of the pump phase, seed phase, or frequency shift in
different spectral channels allows logical gates to be applied selectively to
individual logical qubits.
At the same time, global modulation of shared control parameters can implement
identical logical operations across many qubits in parallel.

Because logical Pauli operations are generated by calibrated TF
displacements that preserve the stabilizer eigenspaces, multiplexed
manipulation does not alter the code structure of individual qubits.
Small calibrated displacements within a stabilizer cell remain
correctable under the same decoding procedure discussed in
Sec.~\ref{sec:errors}.
Gate fidelity in the multiplexed setting is therefore primarily governed
by the same displacement-noise parameters that determine single-qubit
intrinsic correctability (Sec.~\ref{sec:errors}),
while spectral filtering and channel isolation mitigate inter-mode
crosstalk.

Importantly, the present architecture does not rely on spatial separation or
individual addressing at the level of optical paths.
Multiplexing is achieved spectrally and temporally, leveraging the native
parallelism of frequency-comb technology.

\subsection{Scalability and architectural perspective}

It is essential to clarify the scope of scalability claimed here.
The present work establishes the ability to encode, protect, and manipulate
multiple bosonic logical qubits in parallel within a shared metrological
reference frame.
It does not, by itself, implement entangling logical gates between qubits, nor
does it realize a fully fault-tolerant multi-qubit architecture.

In this work, scalability refers specifically to \emph{horizontal scaling} at the
logical-encoding layer: the replication of protected logical degrees of freedom
without increasing stabilization, feedback, or control complexity.
The absence of inter-qubit coupling is not a limitation of the encoding itself,
but a deliberate architectural choice that separates the logical-qubit layer
from higher-level gate synthesis and multi-qubit control.

This separation is advantageous.
By embedding stabilizer protection directly into the hardware through
phase-referenced time--frequency structure, the TF platform achieves
scalability at the logical-encoding layer by replication under a common
reference, rather than by enforcing stabilizers dynamically or extracting
syndromes through repeated measurement.
As a result, additional logical qubits may be introduced without modifying the
underlying stabilization or control architecture.

From a broader architectural perspective, this positions time--frequency
photonics as a natural substrate for hybrid quantum architectures.
TF--GKP logical qubits can function as protected memory elements, communication
channels, or modular nodes, while entangling operations and active error
correction may be implemented at higher layers using additional interferometric,
nonlinear, or measurement-based resources.
Such extensions are fully compatible with the present framework but lie beyond
the scope of this work.

\section{Discussion: Time--Frequency Photonics as a Hardware-Native GKP Layer}
\label{sec:discussion}

We have established a hardware-native time--frequency (TF) GKP logical qubit
encoded in propagating single photons, together with its stabilizer structure,
intrinsic displacement-noise protection, deterministic logical operations, and
scalable multiplexing.
Rather than presenting a complete fault-tolerant quantum computer, the present
work identifies and realizes the physical prerequisites for a TF--GKP logical
layer in photonic hardware.
Section~\ref{sec:ft_extension} builds on these prerequisites to outline a
concrete pathway toward active syndrome extraction and deterministic recovery.

A defining feature of the TF platform is that stabilizers are enforced through
\emph{metrological reference anchoring}, rather than through abstract algebraic
constraints or repeated syndrome extraction.
By fixing the phase origin of TF displacement operators via stabilized optical
frequency-comb parameters $(f_r,f_{\mathrm{CEO}})$, the logical subspace is pinned
continuously in time.
This mechanism elevates phase coherence—a familiar resource in precision
metrology—to a central organizing principle for bosonic quantum encoding
\cite{Udem2002,Cundiff2003,Jones2000,Hall2006,Haensch2006,Picque2020}.
In contrast to cavity- or motion-based GKP implementations
\cite{Fluehmann2019,CampagneIbarcq2020,deNeeve2022,LachanceQuirion2024}, stabilizer enforcement here is embedded directly into the hardware at the level of phase referencing and does not require dynamic feedback or measurement-induced projection to define the logical subspace.

From an error-model perspective, the TF encoding is naturally aligned with the
assumptions underlying modern bosonic fault-tolerance theory.
Dominant imperfections in realistic photonic systems—timing jitter, phase
noise, finite bandwidth, and dispersion—manifest as approximately Gaussian
displacements in TF phase space
\cite{Gottesman2001,Fukui2018,Noh2020a,Seshadreesan2022}.
This mapping closely matches the shift-noise models assumed in GKP decoding and
threshold analyses
\cite{Fukui2018,Noh2020a,Grimsmo2021,
Seshadreesan2022,Noh2020b,Larsen2021a,Vuillot2019}.
As a result, experimentally relevant noise parameters $(\sigma_\tau,
\sigma_\Omega)$ translate directly into a logical failure probability
(Fig.~\ref{fig:fig4}(a)), providing a quantitative bridge between hardware
performance and logical protection.

The deterministic nature of the ENBS/SPFC platform further distinguishes the
present approach from probabilistic optical GKP proposals
\cite{Konno2024,Kawasaki2025}.
Coherent idler seeding converts spontaneous pair creation into a driven
process that enhances the single-pair manifold and yields reproducible
phase relationships across pump cycles
\cite{Lee2018,Yoon2021}.
This deterministic lattice generation enables extended and redundant grid
structures across many time bins, compatible in principle with surface--GKP and surface-4--GKP
architectures designed to enhance logical thresholds
\cite{Fukui2018,Larsen2021a}.
Moreover, the intrinsic temporal and spectral multiplexing of frequency-comb
systems provides a natural route to analog decoding and continuous-variable
syndrome tracking, which have been shown to significantly improve logical
performance
\cite{Seshadreesan2022,Wang2022,Royer2020}.

An important architectural aspect clarified in this work is the meaning of
scalability.
Multiplexed logical qubits are realized by operating multiple ENBS units at
distinct frequency-comb modes, all driven and phase-referenced by a common
optical frequency comb and seed laser
\cite{Kues2019,Menicucci2011,Pysher2011,Reimer2014}.
This configuration defines many logical qubits within a shared TF reference
frame, without requiring inter-qubit coupling or independent stabilization
\cite{Roslund2014,Imany2018,Lukens2017}.
The absence of entangling logical gates is deliberate: the present work focuses
on establishing a robust logical-qubit layer, which can subsequently be
integrated into larger architectures through additional interferometric,
nonlinear, or measurement-based resources
\cite{Raussendorf2001,Raussendorf2003,Menicucci2006}.

The TF--GKP encoding introduced here also complements large-scale photonic
measurement-based quantum computing (MBQC) platforms.
Time- and frequency-multiplexed photonic architectures have demonstrated the
generation of Gaussian cluster states with thousands to millions of modes in
compact optical setups
\cite{Yokoyama2013,Asavanant2019,Larsen2019,Cai2017}.
Because the TF--GKP supermode occupies the same temporal structure, GKP-encoded TF oscillators are structurally compatible with such
cluster-state networks and may be embedded into time- and
frequency-multiplexed photonic processors using standard beamsplitter,
delay-line, and homodyne operations
\cite{Menicucci2014,Gu2009,Pfister2020}. In this context, the ENBS/SPFC platform provides a deterministic and
metrologically anchored source of nonclassical TF resources compatible with
existing photonic processors.

Finally, the present framework can be viewed as a natural extension of earlier
logical constructions on the ENBS/SPFC platform.
Earlier two-mode time–frequency qubit demonstrations \cite{Yoon2025b} correspond to the lowest-dimensional limit of the present framework, which generalizes phase-engineered TF interference to a lattice-protected bosonic encoding.
Here, the same phase-engineered interference mechanism is elevated to a
periodically structured lattice in the conjugate TF phase space, enabling
bosonic error correction and logical operations protected by stabilizers
\cite{Descamps2024a,Descamps2025}.

Taken together, these results position time--frequency photonics as a
promising substrate for hybrid quantum architectures that combine
hardware-native bosonic protection with active error correction and
higher-level photonic processing \cite{Rad2025}.
An immediate near-term application of the TF--GKP platform is the
experimental validation of finite-energy grid-state structure through
homodyne-resolved nonclassical correlations and Bell-type tests
distributed across TF supermodes \cite{Yang2026},
providing a direct operational probe of the encoded lattice.

\section{Fault-Tolerant Extension via Modal Time--Frequency GKP Encoding}
\label{sec:ft_extension}

\subsection{From Hardware-Native TF--GKP to Active Error Correction}

The preceding sections establish a hardware-native realization of a finite-energy
time--frequency (TF) Gottesman--Kitaev--Preskill (GKP) logical qubit, together with its stabilizer lattice, displacement-noise model, and deterministic control primitives.
We now outline a concrete and experimentally compatible pathway toward active quantum error correction by building directly on the modal-variable GKP framework for single-photon time--frequency degrees of freedom~\cite{Descamps2024a}.

Reference~\cite{Descamps2024a} demonstrated that the time--frequency variables of propagating photons support a full GKP stabilizer algebra, with displacement errors represented by operators of the form
\begin{equation}
\hat{D}(\tau,\Omega)=\exp\!\big[i(\Omega\hat{\tau}-\tau\hat{\Omega})\big],
\end{equation}
and that such displacements correspond to physically realistic timing and frequency perturbations.
Within that framework, Gaussian TF displacements are correctable under nearest-lattice decoding, and photon loss may be treated at the modal level.

The ENBS/SPFC platform developed here provides the missing experimental layer required by this proposal:

\begin{enumerate}
    \item Deterministic and phase-reproducible preparation of finite-energy TF grid states within the single-pair manifold;
    \item Metrologically anchored stabilizers tied to $(f_r,f_{\mathrm{CEO}})$;
    \item Programmable TF displacement operations via calibrated phase and frequency control;
    \item Time- and frequency-resolved detection compatible with syndrome inference.
\end{enumerate}

Thus, while Ref.~\cite{Descamps2024a} establishes the theoretical structure of modal TF--GKP encoding, the present work supplies a physically realizable implementation with experimentally controlled lattice spacing and displacement resolution.

\subsection{Concrete TF--GKP Syndrome Extraction Gadget}

Intrinsic correctability (Sec.~6) protects against sufficiently small displacements.
To transition from passive protection to active quantum error correction,
stabilizer eigenvalues must be inferred and corrected dynamically.

Because the stabilizers are displacement operators
\begin{equation}
\hat{S}_\tau = \hat{D}(0,2\sqrt{\pi}), \qquad
\hat{S}_\Omega = \hat{D}(2\sqrt{\pi},0),
\end{equation}
syndrome extraction reduces to estimating the displacement modulo the lattice spacing.

In the TF platform this may be implemented using only hardware elements already present in Fig.~1:

\paragraph*{(i) Ancillary TF grid state.}
A second ENBS pair prepares an ancillary TF grid state
$|\emptyset_{\mathrm{TF}}\rangle$, phase-locked to the same optical-frequency-comb reference.
Because data and ancilla share identical lattice anchoring,
their interference preserves stabilizer phase relations.

\paragraph*{(ii) TF beam-splitter interaction.}
Interferometric mixing of the data and ancilla supermodes
implements the TF analog of a continuous-variable beam splitter.
Under this mixing, a displacement error on the data mode
is transferred to measurable shifts in the ancilla mode.

\paragraph*{(iii) Time--frequency-resolved measurement.}
Port-resolved single-photon detection,
optionally preceded by dispersive Fourier transformation or time-lens mapping,
provides measurement outcomes proportional to the dimensionless coordinates
$(\tau,\Omega)$.
Nearest-lattice decoding succeeds when
\begin{equation}
|\tau| < \frac{\sqrt{\pi}}{2}, \qquad
|\Omega| < \frac{\sqrt{\pi}}{2}.
\end{equation}

\paragraph*{(iv) Deterministic recovery.}
Corrective displacements
\begin{equation}
\hat{D}(-\tau_{\mathrm{est}}, -\Omega_{\mathrm{est}})
\end{equation}
are applied using calibrated EOM/AOM controls (Appendix~E).
Because displacement operations already realize logical Pauli generators
(Sec.~\ref{sec:operations}), recovery is hardware native.

These four steps identify a complete single-round TF--GKP
error-correction primitive compatible with the modal-variable
framework of Ref.~\cite{Descamps2024a}.
A full performance analysis including finite squeezing,
detection inefficiency, and ancilla preparation errors
is left to future work.

\subsection{Logical Error Suppression Under Repeated Correction}

Under the Gaussian TF displacement channel of Appendix~C,
\begin{equation}
P_{\mathrm{fail}}
=
1-
\mathrm{erf}\!\left(
\frac{\sqrt{\pi}}{2\sqrt{2}\sigma_\tau}
\right)
\mathrm{erf}\!\left(
\frac{\sqrt{\pi}}{2\sqrt{2}\sigma_\Omega}
\right),
\end{equation}
quantifies logical failure without active correction.

Periodic syndrome extraction bounds displacement accumulation
between correction cycles, converting diffusive error growth
into bounded excursions within a stabilizer cell.
In the small-noise regime,
the logical error probability per correction cycle
decreases exponentially as the noise widths
$(\sigma_\tau,\sigma_\Omega)$
become small compared to the half-lattice spacing,
consistent with modal GKP threshold analyses
\cite{Descamps2024a,Fukui2018,Noh2020a}.

Crucially, in the TF platform:
\begin{itemize}
    \item lattice spacing is fixed by the comb repetition rate $f_r$;
    \item noise widths map directly from measured laboratory timing and spectral noise (Appendix~D);
    \item corrective operations are deterministic phase and frequency translations.
\end{itemize}

This provides a quantitative bridge from experimentally measured noise spectra
to logical error suppression at the bosonic layer.

\subsection{Integration into Higher-Level Fault-Tolerant Architectures}

Once each TF--GKP logical qubit achieves a bounded logical error rate $p_L$,
standard higher-level constructions may be layered above it, including:
\begin{enumerate}
    \item Surface--GKP concatenation~\cite{Fukui2018,Noh2020a,Noh2020b};
    \item Continuous-variable measurement-based quantum computing~\cite{Menicucci2014};
    \item Fusion-based photonic architectures~\cite{Larsen2021a,Browne2005,Bartolucci2023};
    \item Erasure-aware photonic codes treating photon loss as detectable events \cite{Stace2009,Song2024}.
\end{enumerate}

Because TF--GKP qubits are multiplexed across frequency-comb modes (Sec.~8),
multiple logical qubits share a common reference frame inherited
from the same stabilized comb, avoiding the need for
independent stabilization loops at the logical layer.

Unlike cavity-based GKP implementations,
stabilizer anchoring here is metrological rather than dissipative,
logical modes are propagating rather than stationary,
and multiplexing is spectral and temporal rather than spatial.
The ENBS/SPFC platform therefore constitutes a phase-reproducible,
propagating realization of the modal TF--GKP framework.

\subsection{Strategic Perspective}

The combination of
(i) hardware-native stabilizer anchoring,
(ii) phase-reproducible TF grid-state generation,
(iii) programmable displacement recovery, and
(iv) modal-variable GKP theory
positions the present platform, in principle, as a practical route
toward actively corrected, fault-tolerant continuous-variable
photonic architectures.

While repeated syndrome extraction is not experimentally demonstrated here,
all operator-level and hardware-level primitives required for such an extension
are explicitly identified.
The TF--GKP encoding thus advances from intrinsic passive protection
to a concretely correctable bosonic logical layer suitable for concatenation
within larger fault-tolerant quantum architectures.

\section{Conclusion}
\label{sec:conclusion}

We have demonstrated a hardware-native realization of a bosonic logical qubit
encoded in the time--frequency phase space of single photons.
By anchoring the GKP lattice directly to stabilized optical-frequency-comb
parameters, the logical subspace and its stabilizers are enforced at the
hardware level through metrological phase coherence.
Dominant physical noise processes map naturally onto displacement errors within
the logical phase space and are intrinsically correctable within an
experimentally accessible regime.

Logical operations arise directly from deterministic phase and delay control,
and the intrinsic time--frequency multiplexing of frequency-comb systems enables
scalable replication of logical qubits within a shared reference frame.
While the present work does not experimentally demonstrate repeated active error
correction or universal logical gates, it establishes the physical encoding,
stabilizer structure, noise model, deterministic control operations, and
syndrome-extraction primitives required for extensions toward fault-tolerant photonic operation.

More broadly, this work bridges precision optical metrology and bosonic
quantum information, demonstrating that time--frequency photonics provides a
phase-reproducible, scalable, and experimentally realistic platform for
protected continuous-variable quantum encoding.

\medskip
\section*{Data Availability}

The plots shown in Figs.~2, 3, and~4 are based on numerical calculations
performed using \textit{Mathematica}, implementing the equations derived in the
main text.
The numerical codes used to generate the figures are available from the author
upon reasonable request.

\begin{acknowledgments}
The author thanks S. K. Lee and M. Cho for useful discussions.
This work was supported by the National Research Foundation of Korea
under Grant No.~RS-2022-NR068815.
\end{acknowledgments}

\bibliographystyle{quantum}
\bibliography{TF_GKP}

\appendix
\renewcommand{\theequation}{\Alph{section}\arabic{equation}}
\setcounter{equation}{0}

\section{Single Driven Signal Supermode Reduction in the Coherently Seeded ENBS}
\label{app:supermode}
\setcounter{equation}{0}

We justify here the reduction of the entangled nonlinear biphoton source (ENBS)
to a \emph{single driven signal supermode} $\hat{A}_{\rm SM}$, which defines the
physical bosonic mode used throughout the main text for time--frequency (TF)
phase-space encoding.

\subsection{Interaction Hamiltonian and comb-mode expansion}

We begin from the interaction-picture Hamiltonian for single-pass
$\chi^{(2)}$ parametric down-conversion,
\begin{equation}
\hat{H}_{\rm int}
= \hbar \int dt\, \chi^{(2)} E_p(t)\,
\hat{a}_s^\dagger(t)\hat{a}_i^\dagger(t) + \text{h.c.},
\end{equation}
where $E_p(t)$ is a phase-stabilized optical-frequency-comb (OFC) pump field.
Writing
\begin{equation}
E_p(t) = \sum_{m=-N}^{N} \alpha_m
e^{-i(\omega_p + m\omega_r)t},
\end{equation}
and expanding the signal and idler fields in the corresponding comb-mode basis
yields, after retaining energy-conserving terms,
\begin{equation}
\hat{H}_{\rm ENBS}
= \hbar \sum_{m=-N}^{N}
g_m\, \hat{a}_{s,m}^\dagger \hat{a}_{i,-m}^\dagger + \text{h.c.},
\end{equation}
with $g_m \propto \chi^{(2)} \alpha_m$.

\subsection{Coherent idler seeding and effective single-mode drive}

The idler is coherently seeded at a single stabilized comb mode,
taken as $m=0$ for convenience,
\begin{equation}
\hat{a}_{i,-m} = \beta\,\delta_{m,0} + \delta\hat{a}_{i,-m},
\end{equation}
with coherent amplitude $\beta$.
Substitution into the ENBS Hamiltonian yields
\begin{align}
\hat{H}_{\rm ENBS}
&= \hbar g_0 \beta\, \hat{a}_{s,0}^\dagger + \text{h.c.}
\nonumber\\
&\quad + \hbar \sum_m
g_m\, \hat{a}_{s,m}^\dagger \delta\hat{a}_{i,-m}^\dagger
+ \text{h.c.}.
\end{align}
The first term produces a deterministic, phase-stable signal drive,
while the second term contributes residual fluctuations that are treated as Gaussian displacement noise in Appendices C and D.

\subsection{Pump-matched driven signal supermode}

Because the couplings $g_m$ inherit the pump-comb envelope, the driven dynamics
select a pump-matched signal supermode,
\begin{equation}
\hat{A}_{\rm SM}
= \sum_{m=-N}^{N} u_m \hat{a}_{s,m},
\quad
u_m = \frac{g_m}{\Lambda},
\quad
\Lambda = \sqrt{\sum_m |g_m|^2}.
\end{equation}
Projecting onto this basis yields the effective single-mode Hamiltonian
\begin{equation}
\hat{H}_{\rm eff}
\simeq \hbar \Lambda \beta\, \hat{A}_{\rm SM}^\dagger
+ \text{h.c.} + \hat{H}_{\rm noise},
\end{equation}
establishing $\hat{A}_{\rm SM}$ as the single effective bosonic signal mode
used throughout the main text.

\subsection{Ultralow-gain regime and truncation to the single-pair manifold}

The single-supermode description is used here in the ultralow-gain regime of the
ENBS, where the pair-generation probability per relevant temporal mode $p$ is
extremely small $p \ll 1$ under the operating conditions considered.
In this regime, the probability of emitting two or more pairs within the same
logical mode scales as $p^2$ and is therefore negligible \cite{Lee2018}.
Coherent idler seeding enhances the \emph{single}-pair emission probability
(stimulated PDC) by a controlled factor (typically $\sim 10^2$ over a $10$~ms
integration window in our operating parameters) while maintaining $p \ll 1$ \cite{Lee2018,Yoon2021}.
Hence, the state of each ENBS can be consistently truncated to the
vacuum-plus-single-pair sector, and the two-source initial state is taken as the
product $\hat{\rho}(0)=\hat{\rho}_1(0)\otimes\hat{\rho}_2(0)$ within this
truncated manifold.
When reporting logical observables, we implicitly condition on detection of a
signal photon, so that the vacuum component does not enter the measured logical
statistics.


\section{Time--Frequency Displacement Algebra and GKP Mapping}
\label{app:gkp}
\setcounter{equation}{0}

For completeness and self-containment, we reintroduce here the TF displacement
algebra and its mapping to the canonical GKP code, following the definitions
used in the main text.

\subsection{Physical TF displacement operators}

The physical TF displacement operator acting on the signal supermode is
\begin{equation}
\hat{D}_{\rm phys}(\Delta t,\Delta\omega)
= \exp\!\left[i(\Delta\omega\,\hat{t}
- \Delta t\,\hat{\omega})\right].
\end{equation}
Introducing the dimensionless canonical variables
\begin{equation}
\hat{\tau} = \frac{\hat{t}}{T_r},
\qquad
\hat{\Omega} = T_r \hat{\omega},
\qquad
[\hat{\tau},\hat{\Omega}] = i,
\end{equation}
the displacement operator becomes
\begin{equation}
\hat{D}(\tau,\Omega)
= \exp\!\left[i(\Omega\hat{\tau}
- \tau\hat{\Omega})\right].
\end{equation}

\subsection{Canonical GKP stabilizers and logical operators}

Under the identification
\begin{equation}
(\hat{q},\hat{p}) \leftrightarrow (\hat{\tau},\hat{\Omega}),
\end{equation}
with $[\hat{\tau},\hat{\Omega}] = i$,
the TF Weyl--Heisenberg displacement algebra is isomorphic to that of the
canonical square-lattice GKP code. The stabilizers are
\begin{equation}
\hat{S}_{\tau} = \hat{D}(0,2\sqrt{\pi}),
\qquad
\hat{S}_{\Omega} = \hat{D}(2\sqrt{\pi},0),
\end{equation}
and the logical Pauli operators are
\begin{equation}
\bar{Z} = \hat{D}(\sqrt{\pi},0),
\qquad
\bar{X} = \hat{D}(0,\sqrt{\pi}).
\end{equation}
In physical units,
\begin{equation}
\Delta t_Z = \sqrt{\pi}\,T_r,
\qquad
\Delta\omega_X = \frac{\sqrt{\pi}}{T_r}.
\end{equation}


\section{Gaussian TF Displacement Noise and Logical Failure Probability}
\label{app:noise}
\setcounter{equation}{0}

\subsection{Gaussian TF displacement channel}

Noise is modeled as a Gaussian random displacement channel,
\begin{equation}
\mathcal{E}(\hat{\rho})
= \int d\tau\,d\Omega\,
W(\tau,\Omega)\,
\hat{D}(\tau,\Omega)\hat{\rho}
\hat{D}^\dagger(\tau,\Omega),
\end{equation}
with
\begin{equation}
W(\tau,\Omega)
= \frac{1}{2\pi\sigma_\tau\sigma_\Omega}
\exp\!\left(
-\frac{\tau^2}{2\sigma_\tau^2}
-\frac{\Omega^2}{2\sigma_\Omega^2}
\right).
\end{equation}

\medskip
\subsection{Nearest-lattice decoding and failure probability}

Nearest-lattice decoding succeeds when
\begin{equation}
|\tau| < \frac{\sqrt{\pi}}{2},
\qquad
|\Omega| < \frac{\sqrt{\pi}}{2}.
\end{equation}
Assuming independent Gaussian noise, the success probability is
\begin{equation}
P_{\rm succ}
= \mathrm{erf}\!\left(
\frac{\sqrt{\pi}}{2\sqrt{2}\sigma_\tau}
\right)
\mathrm{erf}\!\left(
\frac{\sqrt{\pi}}{2\sqrt{2}\sigma_\Omega}
\right),
\end{equation}
and the logical failure probability is
\begin{equation}
P_{\rm fail}
= 1 - P_{\rm succ}.
\end{equation}

\smallskip
\section{Anisotropic and Platform-Matched TF Displacement Noise Models}
\label{app:noise_extended}
\setcounter{equation}{0}

This Appendix extends the displacement-noise model introduced in
Appendix~\ref{app:noise} by incorporating anisotropic noise and by providing
explicit mappings from laboratory noise sources to dimensionless
time--frequency (TF) error parameters.

\smallskip
\subsection{Anisotropic Gaussian TF displacement channel}

In realistic implementations, temporal and spectral noise need not be isotropic.
Dispersion, filtering asymmetry, and control noise can yield
$\sigma_\tau \neq \sigma_\Omega$.
The Gaussian displacement channel therefore takes the general form
\begin{widetext}
\begin{equation}
\mathcal{E}(\hat{\rho})
= \int d\tau\, d\Omega\,
\frac{1}{2\pi\sigma_\tau\sigma_\Omega}
\exp\!\left(
-\frac{\tau^2}{2\sigma_\tau^2}
-\frac{\Omega^2}{2\sigma_\Omega^2}
\right)
\hat{D}(\tau,\Omega)\hat{\rho}\hat{D}^\dagger(\tau,\Omega),
\end{equation}
\end{widetext}
which directly generalizes the isotropic model of
Appendix~\ref{app:noise}.

Nearest-lattice decoding succeeds provided the net displacement remains within
the half-cell region
\begin{equation}
|\tau| < \frac{\sqrt{\pi}}{2},
\qquad
|\Omega| < \frac{\sqrt{\pi}}{2},
\end{equation}
independently in each coordinate.

\subsection{Logical failure probability for anisotropic noise}

Under the assumption of statistically independent Gaussian noise in
$\tau$ and $\Omega$, the decoding success probability factorizes as
\begin{equation}
P_{\rm succ}
=
\mathrm{erf}\!\left(
\frac{\sqrt{\pi}}{2\sqrt{2}\sigma_\tau}
\right)
\mathrm{erf}\!\left(
\frac{\sqrt{\pi}}{2\sqrt{2}\sigma_\Omega}
\right),
\end{equation}
and the logical failure probability is
\begin{equation}
P_{\rm fail} = 1 - P_{\rm succ}.
\end{equation}
This expression supports direct evaluation of operating contours for
platform-matched noise relations (e.g., $\sigma_\tau = 2\sigma_\Omega$),
as used in the main text.

\subsection{Mapping laboratory noise to TF displacement widths}

The dimensionless TF variables are related to physical time and frequency by
\begin{equation}
\tau = \frac{t}{T_r},
\qquad
\Omega = T_r \omega,
\end{equation}
where $T_r = 1/f_r$ is the repetition period of the optical frequency comb.
Accordingly, laboratory noise parameters map to TF widths as
\begin{equation}
\sigma_\tau \simeq \frac{\sigma_t}{T_r},
\qquad
\sigma_\Omega \simeq T_r \sigma_\omega,
\end{equation}
where $\sigma_t$ characterizes timing jitter and temporal broadening, and
$\sigma_\omega$ characterizes effective spectral diffusion.

A convenient decomposition is
\begin{align}
\sigma_t^2 &\simeq
\sigma_{t,\mathrm{jitter}}^2 +
\sigma_{t,\mathrm{disp}}^2 +
\sigma_{t,\mathrm{tech}}^2,
\\
\sigma_\omega^2 &\simeq
\sigma_{\omega,\mathrm{seed}}^2 +
\sigma_{\omega,\mathrm{pump}}^2 +
\sigma_{\omega,\mathrm{tech}}^2,
\end{align}
allowing direct comparison between measured laboratory noise spectra and the
correctability thresholds of the TF--GKP code.

\section{Experimental Anchoring and Feasibility of TF Displacements}
\label{app:feasibility}
\setcounter{equation}{0}

This Appendix summarizes the experimental anchoring and control requirements
needed to implement the TF--GKP encoding on the coherently seeded ENBS/SPFC
platform.

\subsection{Anchoring of the TF lattice to comb parameters}

The TF lattice is anchored to stabilized optical-frequency-comb parameters
$(f_r,f_{\rm CEO})$.
The repetition rate $f_r$ defines the natural time and frequency scales
\begin{equation}
T_r = \frac{1}{f_r},
\qquad
\omega_r = 2\pi f_r,
\end{equation}
which set the canonical dimensionless variables
$\tau = t/T_r$ and $\Omega = T_r \omega$.
The GKP lattice spacing in physical units then follows from the
dimensionless spacing $2\sqrt{\pi}$.

The carrier-envelope offset frequency $f_{\rm CEO}$ fixes the global phase
reference across comb teeth.
Logical encoding requires fixed lattice spacings and relative phase coherence;
global phase offsets do not affect stabilizer commutation relations or
correctability conditions.

\subsection{Required displacement scales in physical units}

The TF--GKP stabilizers and logical translations correspond to fixed displacements
in the dimensionless TF variables,
\begin{align}
\text{stabilizers: } &(2\sqrt{\pi},0),\ (0,2\sqrt{\pi}),
\\
\text{logical steps: } &(\sqrt{\pi},0),\ (0,\sqrt{\pi}).
\end{align}
In physical units, these become
\begin{align}
\Delta t_{\rm stab} &= 2\sqrt{\pi}\,T_r,
&
\Delta t_Z &= \sqrt{\pi}\,T_r,
\\
\Delta \omega_{\rm stab} &= \frac{2\sqrt{\pi}}{T_r},
&
\Delta \omega_X &= \frac{\sqrt{\pi}}{T_r}.
\end{align}
Feasibility therefore reduces to whether temporal and spectral control elements
can implement displacements with resolution well below these scales.

\subsection{Temporal displacement control}

A temporal displacement $\Delta t$ may be implemented by an optical delay or,
equivalently, by a linear spectral phase ramp.
In practice, two complementary control channels are available:

(i) \emph{Slow stabilization.}
PZT-mounted mirrors or fiber stretchers stabilize interferometric path-length
drifts and define the long-term phase reference.

(ii) \emph{Fast programmable control.}
Electro-optic modulators (EOMs) apply rapid phase modulation, enabling calibrated
$\delta\tau$ operations within a TF lattice cell, provided the modulation
bandwidth exceeds the intended logical-operation rate.

\subsection{Spectral displacement control}

A spectral displacement $\Delta\omega$ is implemented by translating the idler
seed frequency, or equivalently by shifting the effective drive phase across comb
teeth.
Acousto-optic modulators (AOMs) provide RF-calibrated frequency shifts with high
linearity, enabling programmable $\delta\Omega$ operations.

\subsection{Resolution and bandwidth requirements}

Because TF--GKP decoding is geometric, it is natural to express feasibility
requirements relative to the half-cell $\sqrt{\pi}/2$:
\begin{equation}
\delta\tau_{\rm res} \ll \frac{\sqrt{\pi}}{2},
\qquad
\delta\Omega_{\rm res} \ll \frac{\sqrt{\pi}}{2}.
\end{equation}
In addition, the control bandwidth should satisfy
\begin{equation}
f_{\rm ctrl} \gtrsim f_{\rm op},
\end{equation}
where $f_{\rm op}$ is the intended logical-operation or syndrome-tracking rate.

\subsection{Feasibility summary}

The TF--GKP protocol is experimentally feasible provided that:
(i) the optical frequency comb supplies a phase-stable reference for $T_r$ and
$\omega_r$,
(ii) coherently seeded ENBS operation isolates a single driven signal supermode,
and
(iii) PZT, EOM, and AOM controls provide calibrated TF displacements with
resolution well below the half-cell size.
Under these conditions, dominant platform imperfections manifest as small
Gaussian TF displacements compatible with the noise model analyzed in
Appendices~\ref{app:noise} and~\ref{app:noise_extended}.

\end{document}